\documentclass[aps,showpacs,prl,
reprint,
  ]{revtex4-1}
\usepackage{graphicx}
\usepackage{natbib}
\usepackage{amsmath}
\usepackage{epstopdf}
\usepackage{color}
\usepackage{upgreek}
\usepackage{gensymb}
\begin{document}

\title{Mechanical Spectroscopy of Parametric Amplification in a High-Q Membrane Microresonator}
\author{Shuhui Wu}
\author{Jiteng Sheng$^*$}
\author{Xiaotian Zhang}
\author{Yuelong Wu}
\author{Haibin Wu$^\dag$}
\affiliation{State Key Laboratory of Precision Spectroscopy, East China Normal University, Shanghai 200062, P. R. China}
\date{\today}

\begin{abstract}
We develop a stoichiometric SiN membrane-based optoelectromechanical system, in which the properties of the mechanical resonator can be dynamically controlled by the piezoelectric actuation. The mode splitting is observed in the mechanical response spectra of a phase-sensitive degenerate parametric amplifier. In addition, the quality factor Q of the membrane oscillator in this process is significantly enhanced by more than two orders of magnitude due to the coherent amplification, reaching a Q factor of $\sim$ $3\times10^8$ and a $f\times Q$ product of $\sim$ $8\times10^{14}$ at room temperature. The thermomechanical noise squeezing close to -3 dB limitation is also achieved. This system integrates the electrical, optical and mechanical degrees of freedom without compromising the exceptional material properties of SiN membranes, and can be a useful platform for studying cavity optoelectromechanics.
\end{abstract}


\maketitle
Coupling optical and mechanical degrees of freedom via radiation pressure has attracted great attention owing to the wide variety of applications as well as the fundamental studies on the boundary between quantum and classical physics \cite{kippenberg2008,meyer2014}. Optomechanical interactions have been successfully investigated for the control of mechanical motion and optical processes. A large variety of optomechanical systems have been explored and have achieved tremendous progress in the recent years, including ground-state cooling of mechanical resonators \cite{teufel2011,chan2011,peterson2016}, quantum squeezing of mechanical modes \cite{wollman2015,lecocq2015,pirkkalainen2015}, generation of squeezed light \cite{brooks2012,purdy2013,safavi2013}, and so on.

The outstanding challenge to implement such an optomechanical system is the integration of ultrahigh Q mechanical oscillator and high-finesse optical resonator on the same device. A promising optomechanical platform is introduced at Yale, in which a flexible SiN membrane with exceptional mechanical properties is placed inside a high-finesse optical cavity \cite{thompson2008}. This approach avoids compromising the optical and mechanical features by separating them into two physically distinct objects. So far, such a membrane-in-the-middle configuration has been widely adopted for the study of cavity optomechanics \cite{wilson2009,purdy2012,karuza2012,jockel2015,sawadsky2015,xu2017}.

Electrically controllable optomechanical systems exhibits striking advantages and have been a new field of interest recently \cite{lee2010,o2010quantum,bochmann2013,han2016}. The electrical actuation can usually be much stronger than the photonic coupling through the radiation pressure. Typical optoelectromechanical systems utilize the piezoelectricity or dielectricity of the mechanical oscillator itself.

In this Letter, we develop a SiN membrane-based optoelectromechanical system, in which the electrodes and the mechanical oscillator are physically segregated. Not only can we actuate the membrane oscillation electrically, but more importantly, the stress and the spring constant of the membrane oscillator can also be dynamically modulated via the piezoelectric actuation by directly attaching the substrate of the membrane to a ring piezoelectric actuator. Such a membrane-based optoelectromechanical system offers several advantages compared to other systems: (i) The electrically controllable capacity is integrated in the promising SiN membrane oscillator system without reducing its excellent performance, providing an optical-electrical-mechanical and quantum-compatible device. (ii) Such an electrical channel can also be used within a feedback loop. Cooperating with optomechanical interaction, this offers a platform to study many fascinating effects \cite{cohadon1999,villanueva2011,pontin2014,schafermeier2016,rossi2017}. (iii) The piezoelectric way is more straightforward than the reservoir engineering method, without requiring appropriate geometric and material design \cite{patil2015}. (iv) The buildup of the system does not require complicated nano/micro-fabrication techniques, since all the components used are commercially available.

Parametric resonance is of interest to many fields of physics \cite{landau1976}. Optical parametric amplifier has witnessed its great success in quantum optics, nonlinear optics, laser physics, et al. As a counterpart, mechanical parametric effect has attracted significant attention with the development of fabrication techniques recently \cite{rugar1991,turner1998,carr2000,rhoads2010,karabalin2010,s2011,mahboob2014,leuch2016}. Here we study the degenerate parametric amplification in the membrane-based optoelectromechanical system as a demonstration. We observe the mode splitting in the mechanical response spectra of a phase-sensitive parametric amplifier. In addition, we demonstrate that the quality factor of the membrane oscillator can be significantly enhanced by more than two orders of magnitude due to the coherent amplification, reaching a Q factor of $\sim$ $3\times10^8$ and a $f\times Q$ product of $\sim$ $8\times10^{14}$, which is crucial for ground state cooling and quantum manipulation of an optomechanical system at room temperature. We also investigate the nonlinear effect on the parametric amplification as well as the thermomechanical noise squeezing.

In the experiment we use a 50 nm-thick by 500-$\mu$m-square window of stoichiometric low-pressure chemical vapor deposition SiN membrane supported by a 5$\times$5 $mm^2$ size and 200 $\mu$m-thick silicon frame from NORCADA Inc. The resonant frequencies of the membrane's vibrational modes are $f_{ij}  = \sqrt {\sigma (i^2  + j^2 )/4\rho l^2 }$, where $\sigma \sim$ 0.9 GPa is the tensile stress, $\rho \sim$ 2.7 $g/cm^3$ is the mass density, $l$ is the side length of the square membrane, and $i$, $j$ are the positive integer mode indices. The membrane has typical vibrational frequencies in the MHz range and Q factors in the range of $10^5-10^6$ \cite{zwickl2008,jockel2011,wilson2011,yu2012,villanueva2014,li2016}. Recently, ultrahigh Q factors have been demonstrated in SiN membranes \cite{chakram2014,reinhardt2016,norte2016}. Moreover, the optical absorption of the membrane is ultralow for the near infrared light. These features enable the SiN membrane as an excellent candidate for studying optomechanics.
\begin{figure}[ht]
\includegraphics[width=1\linewidth]{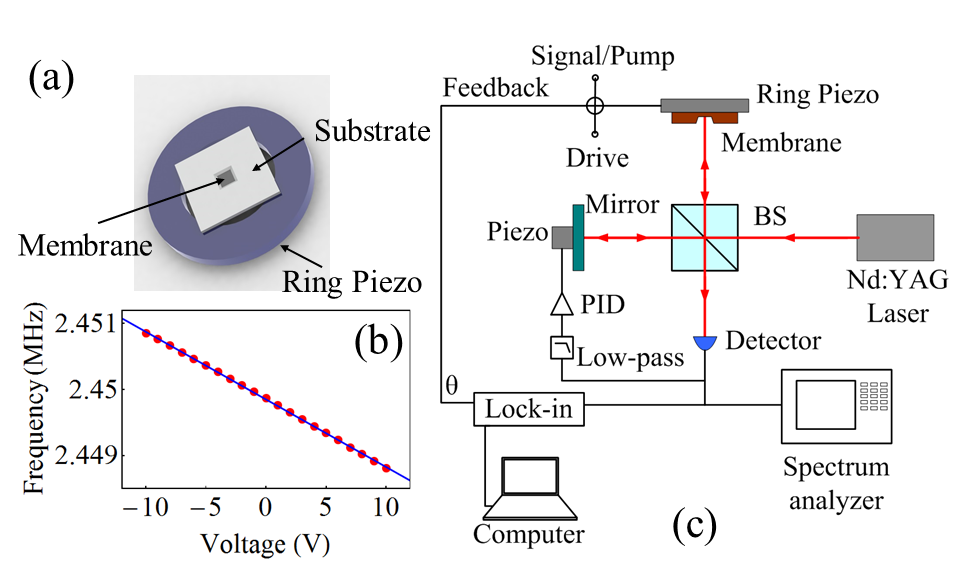}
\caption{(a) The image of the SiN membrane attached to the ring piezo actuator with minimum glue used at three corners. (b) The frequency shift of (3,3) mode as a function of the voltage applied on the ring piezo actuator. (c) The schematic of the experimental system. The signal/pump fields are for the parametric amplification process, and the drive/feedback fields for stabilizing the vibrational eigenmodes. Beam splitter (BS); proportional–integral–derivative controller (PID); the phase channel of the lock-in amplifier ($\theta$). }\label{fig1}
\end{figure}

The membrane chip is directly glued to the ring piezo actuator at three corners, as shown in Fig. 1(a). As the piezo actuator expands in the thickness mode, it will contact the radial mode. Hence, the geometry of the substrate will be perturbed, which leads to the changes of the stress and the spring constant of the membrane. Figure 1(b) presents the dependence of the frequency of (3,3) mode as a function of the voltage applied on the ring piezo actuator. The red dots are the experimental data and the blue line is the linear fit, which indicates that the frequency sensitivity for (3,3) mode is -100Hz/V. We will show that this relation holds for a wide dynamic range and can be utilized to study many fascinating effect, such as parametric amplification. The frequency dependence for several other modes is also investigated. We find that the induced frequency fluctuations of different vibrational modes under such conditions are highly correlated. This feature is utilized to stabilize the vibrational eigenmodes in the experiment, which has a faster response compared to the photothermal feedback \cite{patil2015,gavartin2013}.

The experimental schematic is shown in Fig. 1(c). The membrane is placed in a vacuum chamber which is ion-pumped to $\sim$ 10$^{-8}$ torr. The electric fields applied on the ring piezo actuator include the signal/pump fields for the parametric amplifier, and the drive/feedback fields for stabilizing the vibrational eigenmodes. The mechanical oscillations are transduced optically by a Michelson interferometer with displacement sensitivity better than 10$^{-14}$ m/Hz$^{1/2}$. It is worth noting that the displacement sensitivity can be significantly enhanced by placing the current membrane system inside a high-finesse optical cavity without any modification. The Michelson interferometer is stabilized with a second piezo actuator, as shown in Fig. 1(c).  

The equation of motion for a parametrically excited mechanical oscillator is \cite{landau1976}
\begin{equation}\label{eq1}
\ddot x + \frac{{\omega _0 }}{Q}\dot x + \frac{{k_0 +k_p\cos 2(\omega _0+\Delta _p )t}}{m}x = \frac{{F_0 }}{m}\cos [(\omega _0  + \Delta _s )t + \phi ],
\end{equation}
where x(t) is the membrane displacement, m is the effective mass,  ${\omega _0 }$ is the eigenfrequency of the mechanical oscillator, Q is the quality factor, $k_0  = m\omega _0^2$ is the spring constant, $k_p$($F_0$) is the amplitude of the pump (signal) field, $\Delta _p$($\Delta _s$) is the frequency detuning of the pump (signal) field, and $\phi$ measures the phase of the signal relative to the pump field.

Assume the system is in the limit of weak damping and small oscillation, the secular perturbation theory \cite{lifshitz2010} can be used to solve Eq. (1). We consider the parametrically pumped mechanical response spectrum of the membrane oscillator under the degenerate case, where the pump frequency is always tuned to be twice the signal frequency, i.e. $\Delta _p  = \Delta _s  = \Delta$. Then the response amplitude is 
\begin{equation}\label{eq2}
a =  - \frac{{2\Delta  + i + (k_p /k_{th} )e^{ - 2i\phi } }}{{4\Delta ^2  + 1 - (k_p /k_{th} )^2 }}.
\end{equation}
Here $k_{th}  = 2k_0 /Q$ is the instability threshold of the parametric amplification, and a constant response factor is omitted in Eq. (2). 

\begin{figure}[ht]
\includegraphics[width=1\linewidth]{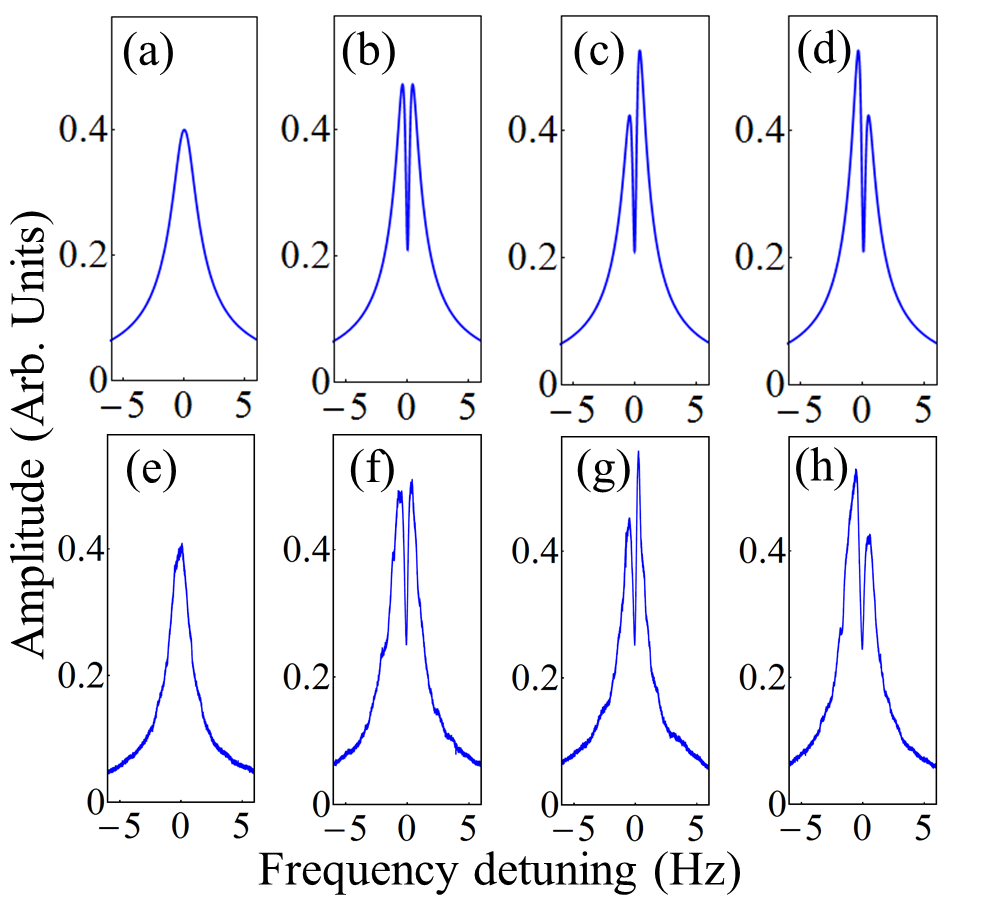}
\caption{Theoretical simulations for (a) without parametric pump; (b) $\phi  = 45^\circ$; (c) $\phi  = 45^\circ  - 1.5^\circ$; (d) $\phi  = 45^\circ  + 1.5^\circ$.  $k_p /k_{th}  = 0.9$ for all the cases in (a-d). (e-f) Experimental measurements for the cases corresponding to (a-d), respectively.}\label{fig2}
\end{figure}

As one can see in Eq. (2), the parametrically pumped mechanical response spectrum is sensitive to the relative phase between the pump and signal fields. Figure 2(a) shows a Lorentzian profile of the signal field when the parametric pump field is absent. When the signal field has a phase $\pi$/4 relative to the pump, i.e. $\phi  = 45^\circ$, there is a symmetric mode splitting in the mechanical response spectrum, as shown in Fig. 2(b). The origin of mode splitting in the spectrum is the destructive interference in cooperation with the dissipation of the membrane oscillator. When the signal field is perfectly on resonant with the eigenmode of the membrane oscillator, i.e., $\Delta  = 0$, the signal and pump fields are exactly out of phase, and they will interfere destructively to produce the deamplification for the signal field. The suppression of the signal field is limited to -3 dB when the pump strength approaches to the parametric instability threshold \cite{rugar1991}. When the signal and pump fields are not exactly out of phase ($\phi  = 45^\circ$), the mode splitting becomes asymmetric, as illustrated in Figs. 2(c) and 2(d), in which the relative phase is 1.5 degrees deviated from $\pi$/4, i.e., $\phi  = 45^\circ  \pm 1.5^\circ$. The Fano-like resonance will appear by further increasing the relative phase. Such a spectroscopic signature of parametric amplification in the out of phase situation has also been observed in the optical parametric amplifier \cite{ma2005}. In analogy to the quantum optical systems, the phase-sensitive mechanical parametric amplifier can be useful for applications such as phase-sensitive switching \cite{sheng2012,khadka2013}, information storage \cite{fiore2011}, logic operations \cite{mahboob2008}, et al. Figures 2(e-f) present the experimental measurements correspond to the cases shown in Figs. 2(a-d) respectively. The parametric pump strength is below threshold with $k_p /k_{th}  = 0.9$ in Fig. 2.

\begin{figure}[htbp]
\includegraphics[width=1.05\linewidth]{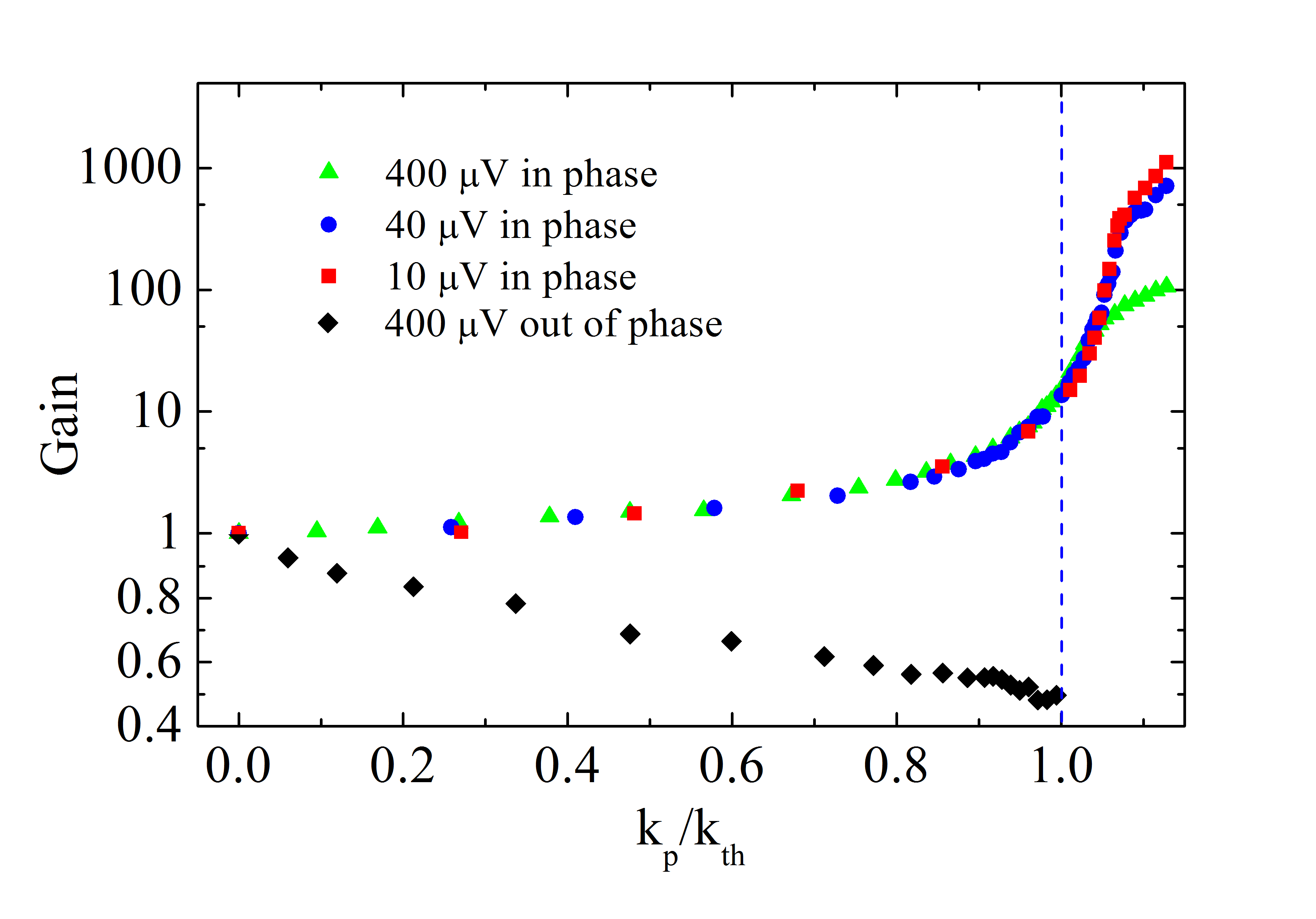}
\caption{Parametric amplification/deamplification as a function of the pump amplitude relative to the pump threshold. The black diamonds show the deamplification case, i.e. $\phi  = \pi /4$. The red squares, blue dots, and green triangles illustrate the parametric amplification ($\phi  = -\pi /4$) for different signal amplitudes respectively. The nonlinearity limits the amplifier gain. The blue dashed line indicates the instability threshold, which is determined by the onset of the self-oscillation.}\label{fig3}
\end{figure}

When the signal field resonates with the eigenmode of the membrane oscillator, and the pump field is perfectly tuned to be twice the signal field, i.e., $\Delta  = 0$, the gain of the parametric amplifier is \cite{lifshitz2010}
\begin{equation}\label{eq3}
G = \frac{{|x|_{pump\;on} }}{{|x|_{pump\;off} }} = [\frac{{\cos ^2 (\phi  + \pi /4)}}{{(1 - k_p /k_{th} )^2 }} + \frac{{\sin ^2 (\phi  + \pi /4)}}{{(1 + k_p /k_{th} )^2 }}]^{1/2}. 
\end{equation}
Figure 3 represents the amplitude gain as a function of the pump amplitude relative to the pump threshold. The black diamonds represent the deamplification case. The sine term in Eq. (3) dominates when $\phi  = \pi /4$, and this leads to the suppression of the signal field, which is limited to -3 dB near the instability threshold. On the contrary, the cosine term remains if $\phi  = -\pi /4$, which results in the amplification of the signal field. Previously, we do not consider the nonlinearity, therefore, as the pump amplitude approaches to the threshold, the parametric gain increases to infinity if $\phi  = -\pi /4$. However, in practical systems the nonlinearity always exists, even comparatively small nonlinearity will limit the amplifier gain. The red squares, blue dots, and green triangles respectively represent the parametric amplification for different signal amplitudes, and they are saturated at different gain levels, as one can see in Fig. 3. The maximum gain obtained below the threshold is $\sim$ 10 instead of infinity. Although the nonlinearity degrades the performance of parametric amplifier, it provides the opportunity that the nonlinear parametric amplifier can be operated above the linear instability threshold, which can be utilized to study non-equilibrium physics \cite{leuch2016}. 

\begin{figure}[htbp]
\includegraphics[width=1\linewidth]{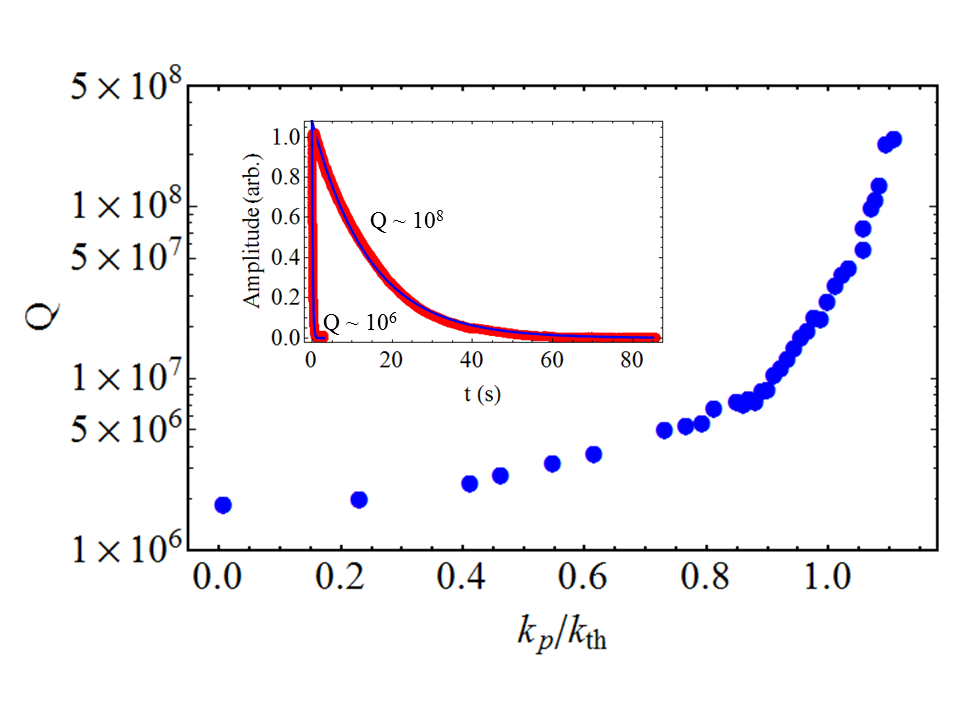}
\caption{The enhanced mechanical Q factor with parametric excitation as a function of the pump amplitude relative to the threshold. The inset is the mechanical ringdown for Q $\sim$ 10$^6$ and 10$^8$. The red thick curves are the experimental data and the blue thin curves are the exponential fits. }\label{fig4}
\end{figure}

A high mechanical Q factor is essential to optomechanics. We find that Q factor with parametric excitation in the current system can be greatly enhanced upon already very high Q in the SiN membrane. Figure 4 shows the dependence of Q factor for different pump amplitudes. The enhancement of Q factor more than two orders of magnitude has been observed due to the coherent amplification, reaching a Q factor of $\sim$ $3\times10^8$ at room temperature. The highest Q factor obtained in the current system is operated slightly above the threshold and in the weak nonlinear region. When the parametric pump is well above the instability threshold, more interesting nonliear dynamic processes can happen, which will be our future studies. The current system offers the opportunity of combining strong nonlinearity and parametric amplification, which can lead to interesting dynamics phenomena. As far as we know, this is the highest Q factor has been directly measured in a SiN membrane. In principle, a higher Q factor is possible to achieve by choosing a vibration mode with a larger intrinsic Q factor. The inset shows the mechanical ringdown measurements correspond to the cases for Q $\sim$ 10$^6$ and 10$^8$.

\begin{figure}[htbp]
\includegraphics[width=1\linewidth]{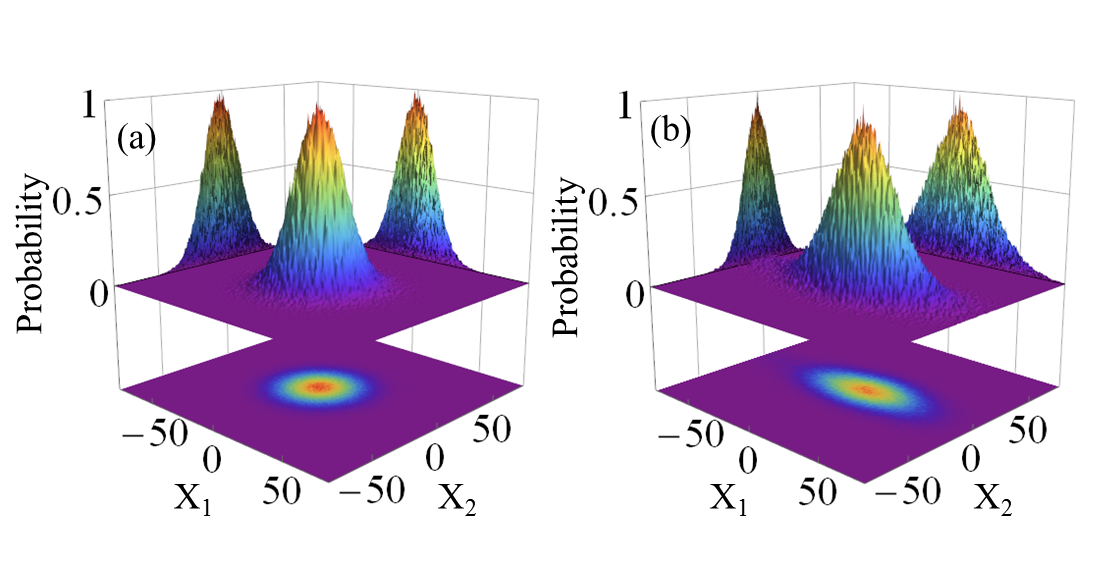}
\caption{The probability density functions of the membrane oscillator in the phase space for the thermal (a) and parametrically squeezed (b) cases, respectively.}\label{fig5}
\end{figure}

A further interesting development related to the parametric amplification is to create a mechanical squeezed state. The thermal motion of the membrane oscillator can be decomposed into two quadratures $x(t) = X_1 (t)\cos \omega _0 t + X_2 (t)\sin \omega _0 t$ in a frame rotating at the mechanical resonance frequency $\omega_0$, where $X_1(t)$ and $X_2(t)$ are stochastic Gaussian noise with variance $< X_1 ^2  >  =  < X_2 ^2  >  = k_B T/m\omega _0^2$, where $k_B$ is the Boltzmann constant and T is the temperature. When the parametric pump field is applied, the quadrature of the motion in phase with the parametric modulation is amplified and the orthogonal quadrature is deamplified, leading to the parametric squeezing of the thermal mechanical motion in the phase space. This can be better understood by substituting the stochastic force for the coherent signal field on the right term of Eq. (1), therefore, the gain of the thermal noise with parametric pumped is sensitive to the relative phase between the pump field and the noise, and we find that the variances of the two quadratures of the motion have a similar behavior of the parametric gain as shown in Fig. 3. $X_1(t)$ and $X_2(t)$ are measured by the two-phase lock-in amplifier (Zurich Instruments HF2LI). The probability density functions of the membrane oscillator in the phase space for the thermal and parametrically squeezed cases are shown in Figs. 5(a) and 5(b), respectively. The noise reduction of -2.1 dB has been observed. A few schemes have been proposed and realized to surpass the -3 dB limit, for example, stabilized parametric squeezing \cite{pontin2014}. 

In conclusion, we have realized a versatile platform for studying cavity optoelectromechanics by combining the piezoelectric effect with a stoichiometric SiN membrane, adding to the capabilities developed in membrane-based optomechanical systems in the last decade. The degenerate parametric amplification is demonstrated in the current system, including mode splitting in the mechanical spectroscopy, Q enhancement, and thermal noise squeezing. The achieved $f\times Q$ product is more than two orders of magnitude greater than the requirements for optomechanical cooling to the ground state from room temperature, i.e. $k_B T_{room} /h$. Moreover, it is possible to squeeze quantum noise of mechanical motion at room temperature and investigate optomechanics in nonlinear quantum regime. This opens the path to applications in fundamental physics, ultrasensitive measurement, and quantum control.

\vskip 0.1in

$^*$jtsheng@lps.ecnu.edu.cn

$^\dag$hbwu@phy.ecun.edu.cn

\bibliographystyle{apsrev4-1}


\begin{thebibliography}{53}%
\makeatletter
\providecommand \@ifxundefined [1]{%
 \@ifx{#1\undefined}
}%
\providecommand \@ifnum [1]{%
 \ifnum #1\expandafter \@firstoftwo
 \else \expandafter \@secondoftwo
 \fi
}%
\providecommand \@ifx [1]{%
 \ifx #1\expandafter \@firstoftwo
 \else \expandafter \@secondoftwo
 \fi
}%
\providecommand \natexlab [1]{#1}%
\providecommand \enquote  [1]{``#1''}%
\providecommand \bibnamefont  [1]{#1}%
\providecommand \bibfnamefont [1]{#1}%
\providecommand \citenamefont [1]{#1}%
\providecommand \href@noop [0]{\@secondoftwo}%
\providecommand \href [0]{\begingroup \@sanitize@url \@href}%
\providecommand \@href[1]{\@@startlink{#1}\@@href}%
\providecommand \@@href[1]{\endgroup#1\@@endlink}%
\providecommand \@sanitize@url [0]{\catcode `\\12\catcode `\$12\catcode
  `\&12\catcode `\#12\catcode `\^12\catcode `\_12\catcode `\%12\relax}%
\providecommand \@@startlink[1]{}%
\providecommand \@@endlink[0]{}%
\providecommand \url  [0]{\begingroup\@sanitize@url \@url }%
\providecommand \@url [1]{\endgroup\@href {#1}{\urlprefix }}%
\providecommand \urlprefix  [0]{URL }%
\providecommand \Eprint [0]{\href }%
\providecommand \doibase [0]{http://dx.doi.org/}%
\providecommand \selectlanguage [0]{\@gobble}%
\providecommand \bibinfo  [0]{\@secondoftwo}%
\providecommand \bibfield  [0]{\@secondoftwo}%
\providecommand \translation [1]{[#1]}%
\providecommand \BibitemOpen [0]{}%
\providecommand \bibitemStop [0]{}%
\providecommand \bibitemNoStop [0]{.\EOS\space}%
\providecommand \EOS [0]{\spacefactor3000\relax}%
\providecommand \BibitemShut  [1]{\csname bibitem#1\endcsname}%
\let\auto@bib@innerbib\@empty
\bibitem [{\citenamefont {Kippenberg}\ and\ \citenamefont
  {Vahala}(2008)}]{kippenberg2008}%
  \BibitemOpen
  \bibfield  {author} {\bibinfo {author} {\bibfnamefont {T.~J.}\ \bibnamefont
  {Kippenberg}}\ and\ \bibinfo {author} {\bibfnamefont {K.~J.}\ \bibnamefont
  {Vahala}},\ }\href@noop {} {\bibfield  {journal} {\bibinfo  {journal}
  {Science}\ }\textbf {\bibinfo {volume} {321}},\ \bibinfo {pages} {1172}
  (\bibinfo {year} {2008})}\BibitemShut {NoStop}%
\bibitem [{\citenamefont {Aspelmeyer}\ \emph {et~al.}(2014)\citenamefont
  {Aspelmeyer}, \citenamefont {Kippenberg},\ and\ \citenamefont
  {Marquardt}}]{meyer2014}%
  \BibitemOpen
  \bibfield  {author} {\bibinfo {author} {\bibfnamefont {M.}~\bibnamefont
  {Aspelmeyer}}, \bibinfo {author} {\bibfnamefont {T.~J.}\ \bibnamefont
  {Kippenberg}}, \ and\ \bibinfo {author} {\bibfnamefont {F.}~\bibnamefont
  {Marquardt}},\ }\href@noop {} {\bibfield  {journal} {\bibinfo  {journal}
  {Reviews of Modern Physics}\ }\textbf {\bibinfo {volume} {86}},\ \bibinfo
  {pages} {1391} (\bibinfo {year} {2014})}\BibitemShut {NoStop}%
\bibitem [{\citenamefont {Teufel}\ \emph {et~al.}(2011)\citenamefont {Teufel},
  \citenamefont {Donner}, \citenamefont {Li}, \citenamefont {Harlow},
  \citenamefont {Allman}, \citenamefont {Cicak}, \citenamefont {Sirois},
  \citenamefont {Whittaker}, \citenamefont {Lehnert},\ and\ \citenamefont
  {Simmonds}}]{teufel2011}%
  \BibitemOpen
  \bibfield  {author} {\bibinfo {author} {\bibfnamefont {J.}~\bibnamefont
  {Teufel}}, \bibinfo {author} {\bibfnamefont {T.}~\bibnamefont {Donner}},
  \bibinfo {author} {\bibfnamefont {D.}~\bibnamefont {Li}}, \bibinfo {author}
  {\bibfnamefont {J.}~\bibnamefont {Harlow}}, \bibinfo {author} {\bibfnamefont
  {M.}~\bibnamefont {Allman}}, \bibinfo {author} {\bibfnamefont
  {K.}~\bibnamefont {Cicak}}, \bibinfo {author} {\bibfnamefont
  {A.}~\bibnamefont {Sirois}}, \bibinfo {author} {\bibfnamefont {J.~D.}\
  \bibnamefont {Whittaker}}, \bibinfo {author} {\bibfnamefont {K.}~\bibnamefont
  {Lehnert}}, \ and\ \bibinfo {author} {\bibfnamefont {R.~W.}\ \bibnamefont
  {Simmonds}},\ }\href@noop {} {\bibfield  {journal} {\bibinfo  {journal}
  {Nature}\ }\textbf {\bibinfo {volume} {475}},\ \bibinfo {pages} {359}
  (\bibinfo {year} {2011})}\BibitemShut {NoStop}%
\bibitem [{\citenamefont {Chan}\ \emph {et~al.}(2011)\citenamefont {Chan},
  \citenamefont {Alegre}, \citenamefont {Safavi-Naeini}, \citenamefont {Hill},
  \citenamefont {Krause}, \citenamefont {Gr{\"o}blacher}, \citenamefont
  {Aspelmeyer},\ and\ \citenamefont {Painter}}]{chan2011}%
  \BibitemOpen
  \bibfield  {author} {\bibinfo {author} {\bibfnamefont {J.}~\bibnamefont
  {Chan}}, \bibinfo {author} {\bibfnamefont {T.~M.}\ \bibnamefont {Alegre}},
  \bibinfo {author} {\bibfnamefont {A.~H.}\ \bibnamefont {Safavi-Naeini}},
  \bibinfo {author} {\bibfnamefont {J.~T.}\ \bibnamefont {Hill}}, \bibinfo
  {author} {\bibfnamefont {A.}~\bibnamefont {Krause}}, \bibinfo {author}
  {\bibfnamefont {S.}~\bibnamefont {Gr{\"o}blacher}}, \bibinfo {author}
  {\bibfnamefont {M.}~\bibnamefont {Aspelmeyer}}, \ and\ \bibinfo {author}
  {\bibfnamefont {O.}~\bibnamefont {Painter}},\ }\href@noop {} {\bibfield
  {journal} {\bibinfo  {journal} {Nature}\ }\textbf {\bibinfo {volume} {478}},\
  \bibinfo {pages} {89} (\bibinfo {year} {2011})}\BibitemShut {NoStop}%
\bibitem [{\citenamefont {Peterson}\ \emph {et~al.}(2016)\citenamefont
  {Peterson}, \citenamefont {Purdy}, \citenamefont {Kampel}, \citenamefont
  {Andrews}, \citenamefont {Yu}, \citenamefont {Lehnert},\ and\ \citenamefont
  {Regal}}]{peterson2016}%
  \BibitemOpen
  \bibfield  {author} {\bibinfo {author} {\bibfnamefont {R.}~\bibnamefont
  {Peterson}}, \bibinfo {author} {\bibfnamefont {T.}~\bibnamefont {Purdy}},
  \bibinfo {author} {\bibfnamefont {N.}~\bibnamefont {Kampel}}, \bibinfo
  {author} {\bibfnamefont {R.}~\bibnamefont {Andrews}}, \bibinfo {author}
  {\bibfnamefont {P.-L.}\ \bibnamefont {Yu}}, \bibinfo {author} {\bibfnamefont
  {K.}~\bibnamefont {Lehnert}}, \ and\ \bibinfo {author} {\bibfnamefont
  {C.}~\bibnamefont {Regal}},\ }\href@noop {} {\bibfield  {journal} {\bibinfo
  {journal} {Physical review letters}\ }\textbf {\bibinfo {volume} {116}},\
  \bibinfo {pages} {063601} (\bibinfo {year} {2016})}\BibitemShut {NoStop}%
\bibitem [{\citenamefont {Wollman}\ \emph {et~al.}(2015)\citenamefont
  {Wollman}, \citenamefont {Lei}, \citenamefont {Weinstein}, \citenamefont
  {Suh}, \citenamefont {Kronwald}, \citenamefont {Marquardt}, \citenamefont
  {Clerk},\ and\ \citenamefont {Schwab}}]{wollman2015}%
  \BibitemOpen
  \bibfield  {author} {\bibinfo {author} {\bibfnamefont {E.~E.}\ \bibnamefont
  {Wollman}}, \bibinfo {author} {\bibfnamefont {C.}~\bibnamefont {Lei}},
  \bibinfo {author} {\bibfnamefont {A.}~\bibnamefont {Weinstein}}, \bibinfo
  {author} {\bibfnamefont {J.}~\bibnamefont {Suh}}, \bibinfo {author}
  {\bibfnamefont {A.}~\bibnamefont {Kronwald}}, \bibinfo {author}
  {\bibfnamefont {F.}~\bibnamefont {Marquardt}}, \bibinfo {author}
  {\bibfnamefont {A.}~\bibnamefont {Clerk}}, \ and\ \bibinfo {author}
  {\bibfnamefont {K.}~\bibnamefont {Schwab}},\ }\href@noop {} {\bibfield
  {journal} {\bibinfo  {journal} {Science}\ }\textbf {\bibinfo {volume}
  {349}},\ \bibinfo {pages} {952} (\bibinfo {year} {2015})}\BibitemShut
  {NoStop}%
\bibitem [{\citenamefont {Lecocq}\ \emph {et~al.}(2015)\citenamefont {Lecocq},
  \citenamefont {Clark}, \citenamefont {Simmonds}, \citenamefont {Aumentado},\
  and\ \citenamefont {Teufel}}]{lecocq2015}%
  \BibitemOpen
  \bibfield  {author} {\bibinfo {author} {\bibfnamefont {F.}~\bibnamefont
  {Lecocq}}, \bibinfo {author} {\bibfnamefont {J.~B.}\ \bibnamefont {Clark}},
  \bibinfo {author} {\bibfnamefont {R.~W.}\ \bibnamefont {Simmonds}}, \bibinfo
  {author} {\bibfnamefont {J.}~\bibnamefont {Aumentado}}, \ and\ \bibinfo
  {author} {\bibfnamefont {J.~D.}\ \bibnamefont {Teufel}},\ }\href@noop {}
  {\bibfield  {journal} {\bibinfo  {journal} {Physical Review X}\ }\textbf
  {\bibinfo {volume} {5}},\ \bibinfo {pages} {041037} (\bibinfo {year}
  {2015})}\BibitemShut {NoStop}%
\bibitem [{\citenamefont {Pirkkalainen}\ \emph {et~al.}(2015)\citenamefont
  {Pirkkalainen}, \citenamefont {Damsk{\"a}gg}, \citenamefont {Brandt},
  \citenamefont {Massel},\ and\ \citenamefont
  {Sillanp{\"a}{\"a}}}]{pirkkalainen2015}%
  \BibitemOpen
  \bibfield  {author} {\bibinfo {author} {\bibfnamefont {J.-M.}\ \bibnamefont
  {Pirkkalainen}}, \bibinfo {author} {\bibfnamefont {E.}~\bibnamefont
  {Damsk{\"a}gg}}, \bibinfo {author} {\bibfnamefont {M.}~\bibnamefont
  {Brandt}}, \bibinfo {author} {\bibfnamefont {F.}~\bibnamefont {Massel}}, \
  and\ \bibinfo {author} {\bibfnamefont {M.~A.}\ \bibnamefont
  {Sillanp{\"a}{\"a}}},\ }\href@noop {} {\bibfield  {journal} {\bibinfo
  {journal} {Physical review letters}\ }\textbf {\bibinfo {volume} {115}},\
  \bibinfo {pages} {243601} (\bibinfo {year} {2015})}\BibitemShut {NoStop}%
\bibitem [{\citenamefont {Brooks}\ \emph {et~al.}(2012)\citenamefont {Brooks},
  \citenamefont {Botter}, \citenamefont {Schreppler}, \citenamefont {Purdy},
  \citenamefont {Brahms},\ and\ \citenamefont {Stamper-Kurn}}]{brooks2012}%
  \BibitemOpen
  \bibfield  {author} {\bibinfo {author} {\bibfnamefont {D.~W.}\ \bibnamefont
  {Brooks}}, \bibinfo {author} {\bibfnamefont {T.}~\bibnamefont {Botter}},
  \bibinfo {author} {\bibfnamefont {S.}~\bibnamefont {Schreppler}}, \bibinfo
  {author} {\bibfnamefont {T.~P.}\ \bibnamefont {Purdy}}, \bibinfo {author}
  {\bibfnamefont {N.}~\bibnamefont {Brahms}}, \ and\ \bibinfo {author}
  {\bibfnamefont {D.~M.}\ \bibnamefont {Stamper-Kurn}},\ }\href@noop {}
  {\bibfield  {journal} {\bibinfo  {journal} {Nature}\ }\textbf {\bibinfo
  {volume} {488}},\ \bibinfo {pages} {476} (\bibinfo {year}
  {2012})}\BibitemShut {NoStop}%
\bibitem [{\citenamefont {Purdy}\ \emph {et~al.}(2013)\citenamefont {Purdy},
  \citenamefont {Yu}, \citenamefont {Peterson}, \citenamefont {Kampel},\ and\
  \citenamefont {Regal}}]{purdy2013}%
  \BibitemOpen
  \bibfield  {author} {\bibinfo {author} {\bibfnamefont {T.}~\bibnamefont
  {Purdy}}, \bibinfo {author} {\bibfnamefont {P.-L.}\ \bibnamefont {Yu}},
  \bibinfo {author} {\bibfnamefont {R.}~\bibnamefont {Peterson}}, \bibinfo
  {author} {\bibfnamefont {N.}~\bibnamefont {Kampel}}, \ and\ \bibinfo {author}
  {\bibfnamefont {C.}~\bibnamefont {Regal}},\ }\href@noop {} {\bibfield
  {journal} {\bibinfo  {journal} {Physical Review X}\ }\textbf {\bibinfo
  {volume} {3}},\ \bibinfo {pages} {031012} (\bibinfo {year}
  {2013})}\BibitemShut {NoStop}%
\bibitem [{\citenamefont {Safavi-Naeini}\ \emph {et~al.}(2013)\citenamefont
  {Safavi-Naeini}, \citenamefont {Gr{\"o}blacher}, \citenamefont {Hill},
  \citenamefont {Chan}, \citenamefont {Aspelmeyer},\ and\ \citenamefont
  {Painter}}]{safavi2013}%
  \BibitemOpen
  \bibfield  {author} {\bibinfo {author} {\bibfnamefont {A.~H.}\ \bibnamefont
  {Safavi-Naeini}}, \bibinfo {author} {\bibfnamefont {S.}~\bibnamefont
  {Gr{\"o}blacher}}, \bibinfo {author} {\bibfnamefont {J.~T.}\ \bibnamefont
  {Hill}}, \bibinfo {author} {\bibfnamefont {J.}~\bibnamefont {Chan}}, \bibinfo
  {author} {\bibfnamefont {M.}~\bibnamefont {Aspelmeyer}}, \ and\ \bibinfo
  {author} {\bibfnamefont {O.}~\bibnamefont {Painter}},\ }\href@noop {}
  {\bibfield  {journal} {\bibinfo  {journal} {Nature}\ }\textbf {\bibinfo
  {volume} {500}},\ \bibinfo {pages} {185} (\bibinfo {year}
  {2013})}\BibitemShut {NoStop}%
\bibitem [{\citenamefont {Thompson}\ \emph {et~al.}(2008)\citenamefont
  {Thompson}, \citenamefont {Zwickl}, \citenamefont {Jayich}, \citenamefont
  {Marquardt}, \citenamefont {Girvin},\ and\ \citenamefont
  {Harris}}]{thompson2008}%
  \BibitemOpen
  \bibfield  {author} {\bibinfo {author} {\bibfnamefont {J.}~\bibnamefont
  {Thompson}}, \bibinfo {author} {\bibfnamefont {B.}~\bibnamefont {Zwickl}},
  \bibinfo {author} {\bibfnamefont {A.}~\bibnamefont {Jayich}}, \bibinfo
  {author} {\bibfnamefont {F.}~\bibnamefont {Marquardt}}, \bibinfo {author}
  {\bibfnamefont {S.}~\bibnamefont {Girvin}}, \ and\ \bibinfo {author}
  {\bibfnamefont {J.}~\bibnamefont {Harris}},\ }\href@noop {} {\bibfield
  {journal} {\bibinfo  {journal} {Nature}\ }\textbf {\bibinfo {volume} {452}},\
  \bibinfo {pages} {72} (\bibinfo {year} {2008})}\BibitemShut {NoStop}%
\bibitem [{\citenamefont {Wilson}\ \emph {et~al.}(2009)\citenamefont {Wilson},
  \citenamefont {Regal}, \citenamefont {Papp},\ and\ \citenamefont
  {Kimble}}]{wilson2009}%
  \BibitemOpen
  \bibfield  {author} {\bibinfo {author} {\bibfnamefont {D.}~\bibnamefont
  {Wilson}}, \bibinfo {author} {\bibfnamefont {C.}~\bibnamefont {Regal}},
  \bibinfo {author} {\bibfnamefont {S.}~\bibnamefont {Papp}}, \ and\ \bibinfo
  {author} {\bibfnamefont {H.}~\bibnamefont {Kimble}},\ }\href@noop {}
  {\bibfield  {journal} {\bibinfo  {journal} {Physical review letters}\
  }\textbf {\bibinfo {volume} {103}},\ \bibinfo {pages} {207204} (\bibinfo
  {year} {2009})}\BibitemShut {NoStop}%
\bibitem [{\citenamefont {Purdy}\ \emph {et~al.}(2012)\citenamefont {Purdy},
  \citenamefont {Peterson}, \citenamefont {Yu},\ and\ \citenamefont
  {Regal}}]{purdy2012}%
  \BibitemOpen
  \bibfield  {author} {\bibinfo {author} {\bibfnamefont {T.}~\bibnamefont
  {Purdy}}, \bibinfo {author} {\bibfnamefont {R.}~\bibnamefont {Peterson}},
  \bibinfo {author} {\bibfnamefont {P.}~\bibnamefont {Yu}}, \ and\ \bibinfo
  {author} {\bibfnamefont {C.}~\bibnamefont {Regal}},\ }\href@noop {}
  {\bibfield  {journal} {\bibinfo  {journal} {New Journal of Physics}\ }\textbf
  {\bibinfo {volume} {14}},\ \bibinfo {pages} {115021} (\bibinfo {year}
  {2012})}\BibitemShut {NoStop}%
\bibitem [{\citenamefont {Karuza}\ \emph {et~al.}(2012)\citenamefont {Karuza},
  \citenamefont {Molinelli}, \citenamefont {Galassi}, \citenamefont
  {Biancofiore}, \citenamefont {Natali}, \citenamefont {Tombesi}, \citenamefont
  {Di~Giuseppe},\ and\ \citenamefont {Vitali}}]{karuza2012}%
  \BibitemOpen
  \bibfield  {author} {\bibinfo {author} {\bibfnamefont {M.}~\bibnamefont
  {Karuza}}, \bibinfo {author} {\bibfnamefont {C.}~\bibnamefont {Molinelli}},
  \bibinfo {author} {\bibfnamefont {M.}~\bibnamefont {Galassi}}, \bibinfo
  {author} {\bibfnamefont {C.}~\bibnamefont {Biancofiore}}, \bibinfo {author}
  {\bibfnamefont {R.}~\bibnamefont {Natali}}, \bibinfo {author} {\bibfnamefont
  {P.}~\bibnamefont {Tombesi}}, \bibinfo {author} {\bibfnamefont
  {G.}~\bibnamefont {Di~Giuseppe}}, \ and\ \bibinfo {author} {\bibfnamefont
  {D.}~\bibnamefont {Vitali}},\ }\href@noop {} {\bibfield  {journal} {\bibinfo
  {journal} {New Journal of Physics}\ }\textbf {\bibinfo {volume} {14}},\
  \bibinfo {pages} {095015} (\bibinfo {year} {2012})}\BibitemShut {NoStop}%
\bibitem [{\citenamefont {J{\"o}ckel}\ \emph {et~al.}(2015)\citenamefont
  {J{\"o}ckel}, \citenamefont {Faber}, \citenamefont {Kampschulte},
  \citenamefont {Korppi}, \citenamefont {Rakher},\ and\ \citenamefont
  {Treutlein}}]{jockel2015}%
  \BibitemOpen
  \bibfield  {author} {\bibinfo {author} {\bibfnamefont {A.}~\bibnamefont
  {J{\"o}ckel}}, \bibinfo {author} {\bibfnamefont {A.}~\bibnamefont {Faber}},
  \bibinfo {author} {\bibfnamefont {T.}~\bibnamefont {Kampschulte}}, \bibinfo
  {author} {\bibfnamefont {M.}~\bibnamefont {Korppi}}, \bibinfo {author}
  {\bibfnamefont {M.~T.}\ \bibnamefont {Rakher}}, \ and\ \bibinfo {author}
  {\bibfnamefont {P.}~\bibnamefont {Treutlein}},\ }\href@noop {} {\bibfield
  {journal} {\bibinfo  {journal} {Nature nanotechnology}\ }\textbf {\bibinfo
  {volume} {10}},\ \bibinfo {pages} {55} (\bibinfo {year} {2015})}\BibitemShut
  {NoStop}%
\bibitem [{\citenamefont {Sawadsky}\ \emph {et~al.}(2015)\citenamefont
  {Sawadsky}, \citenamefont {Kaufer}, \citenamefont {Nia}, \citenamefont
  {Tarabrin}, \citenamefont {Khalili}, \citenamefont {Hammerer},\ and\
  \citenamefont {Schnabel}}]{sawadsky2015}%
  \BibitemOpen
  \bibfield  {author} {\bibinfo {author} {\bibfnamefont {A.}~\bibnamefont
  {Sawadsky}}, \bibinfo {author} {\bibfnamefont {H.}~\bibnamefont {Kaufer}},
  \bibinfo {author} {\bibfnamefont {R.~M.}\ \bibnamefont {Nia}}, \bibinfo
  {author} {\bibfnamefont {S.~P.}\ \bibnamefont {Tarabrin}}, \bibinfo {author}
  {\bibfnamefont {F.~Y.}\ \bibnamefont {Khalili}}, \bibinfo {author}
  {\bibfnamefont {K.}~\bibnamefont {Hammerer}}, \ and\ \bibinfo {author}
  {\bibfnamefont {R.}~\bibnamefont {Schnabel}},\ }\href@noop {} {\bibfield
  {journal} {\bibinfo  {journal} {Physical review letters}\ }\textbf {\bibinfo
  {volume} {114}},\ \bibinfo {pages} {043601} (\bibinfo {year}
  {2015})}\BibitemShut {NoStop}%
\bibitem [{\citenamefont {Xu}\ \emph {et~al.}(2017)\citenamefont {Xu},
  \citenamefont {Kemiktarak}, \citenamefont {Fan}, \citenamefont {Ragole},
  \citenamefont {Lawall},\ and\ \citenamefont {Taylor}}]{xu2017}%
  \BibitemOpen
  \bibfield  {author} {\bibinfo {author} {\bibfnamefont {H.}~\bibnamefont
  {Xu}}, \bibinfo {author} {\bibfnamefont {U.}~\bibnamefont {Kemiktarak}},
  \bibinfo {author} {\bibfnamefont {J.}~\bibnamefont {Fan}}, \bibinfo {author}
  {\bibfnamefont {S.}~\bibnamefont {Ragole}}, \bibinfo {author} {\bibfnamefont
  {J.}~\bibnamefont {Lawall}}, \ and\ \bibinfo {author} {\bibfnamefont
  {J.}~\bibnamefont {Taylor}},\ }\href@noop {} {\bibfield  {journal} {\bibinfo
  {journal} {Nature Communications}\ }\textbf {\bibinfo {volume} {8}},\
  \bibinfo {pages} {14481} (\bibinfo {year} {2017})}\BibitemShut {NoStop}%
\bibitem [{\citenamefont {Lee}\ \emph {et~al.}(2010)\citenamefont {Lee},
  \citenamefont {McRae}, \citenamefont {Harris}, \citenamefont {Knittel},\ and\
  \citenamefont {Bowen}}]{lee2010}%
  \BibitemOpen
  \bibfield  {author} {\bibinfo {author} {\bibfnamefont {K.~H.}\ \bibnamefont
  {Lee}}, \bibinfo {author} {\bibfnamefont {T.~G.}\ \bibnamefont {McRae}},
  \bibinfo {author} {\bibfnamefont {G.~I.}\ \bibnamefont {Harris}}, \bibinfo
  {author} {\bibfnamefont {J.}~\bibnamefont {Knittel}}, \ and\ \bibinfo
  {author} {\bibfnamefont {W.~P.}\ \bibnamefont {Bowen}},\ }\href@noop {}
  {\bibfield  {journal} {\bibinfo  {journal} {Physical review letters}\
  }\textbf {\bibinfo {volume} {104}},\ \bibinfo {pages} {123604} (\bibinfo
  {year} {2010})}\BibitemShut {NoStop}%
\bibitem [{\citenamefont {O’Connell}\ \emph {et~al.}(2010)\citenamefont
  {O’Connell}, \citenamefont {Hofheinz}, \citenamefont {Ansmann},
  \citenamefont {Bialczak}, \citenamefont {Lenander}, \citenamefont {Lucero},
  \citenamefont {Neeley}, \citenamefont {Sank}, \citenamefont {Wang},
  \citenamefont {Weides} \emph {et~al.}}]{o2010quantum}%
  \BibitemOpen
  \bibfield  {author} {\bibinfo {author} {\bibfnamefont {A.~D.}\ \bibnamefont
  {O’Connell}}, \bibinfo {author} {\bibfnamefont {M.}~\bibnamefont
  {Hofheinz}}, \bibinfo {author} {\bibfnamefont {M.}~\bibnamefont {Ansmann}},
  \bibinfo {author} {\bibfnamefont {R.~C.}\ \bibnamefont {Bialczak}}, \bibinfo
  {author} {\bibfnamefont {M.}~\bibnamefont {Lenander}}, \bibinfo {author}
  {\bibfnamefont {E.}~\bibnamefont {Lucero}}, \bibinfo {author} {\bibfnamefont
  {M.}~\bibnamefont {Neeley}}, \bibinfo {author} {\bibfnamefont
  {D.}~\bibnamefont {Sank}}, \bibinfo {author} {\bibfnamefont {H.}~\bibnamefont
  {Wang}}, \bibinfo {author} {\bibfnamefont {M.}~\bibnamefont {Weides}},  \emph
  {et~al.},\ }\href@noop {} {\bibfield  {journal} {\bibinfo  {journal}
  {Nature}\ }\textbf {\bibinfo {volume} {464}},\ \bibinfo {pages} {697}
  (\bibinfo {year} {2010})}\BibitemShut {NoStop}%
\bibitem [{\citenamefont {Bochmann}\ \emph {et~al.}(2013)\citenamefont
  {Bochmann}, \citenamefont {Vainsencher}, \citenamefont {Awschalom},\ and\
  \citenamefont {Cleland}}]{bochmann2013}%
  \BibitemOpen
  \bibfield  {author} {\bibinfo {author} {\bibfnamefont {J.}~\bibnamefont
  {Bochmann}}, \bibinfo {author} {\bibfnamefont {A.}~\bibnamefont
  {Vainsencher}}, \bibinfo {author} {\bibfnamefont {D.~D.}\ \bibnamefont
  {Awschalom}}, \ and\ \bibinfo {author} {\bibfnamefont {A.~N.}\ \bibnamefont
  {Cleland}},\ }\href@noop {} {\bibfield  {journal} {\bibinfo  {journal}
  {Nature Physics}\ }\textbf {\bibinfo {volume} {9}},\ \bibinfo {pages} {712}
  (\bibinfo {year} {2013})}\BibitemShut {NoStop}%
\bibitem [{\citenamefont {Han}\ \emph {et~al.}(2016)\citenamefont {Han},
  \citenamefont {Zou},\ and\ \citenamefont {Tang}}]{han2016}%
  \BibitemOpen
  \bibfield  {author} {\bibinfo {author} {\bibfnamefont {X.}~\bibnamefont
  {Han}}, \bibinfo {author} {\bibfnamefont {C.-L.}\ \bibnamefont {Zou}}, \ and\
  \bibinfo {author} {\bibfnamefont {H.~X.}\ \bibnamefont {Tang}},\ }\href@noop
  {} {\bibfield  {journal} {\bibinfo  {journal} {Physical review letters}\
  }\textbf {\bibinfo {volume} {117}},\ \bibinfo {pages} {123603} (\bibinfo
  {year} {2016})}\BibitemShut {NoStop}%
\bibitem [{\citenamefont {Cohadon}\ \emph {et~al.}(1999)\citenamefont
  {Cohadon}, \citenamefont {Heidmann},\ and\ \citenamefont
  {Pinard}}]{cohadon1999}%
  \BibitemOpen
  \bibfield  {author} {\bibinfo {author} {\bibfnamefont {P.-F.}\ \bibnamefont
  {Cohadon}}, \bibinfo {author} {\bibfnamefont {A.}~\bibnamefont {Heidmann}}, \
  and\ \bibinfo {author} {\bibfnamefont {M.}~\bibnamefont {Pinard}},\
  }\href@noop {} {\bibfield  {journal} {\bibinfo  {journal} {Physical Review
  Letters}\ }\textbf {\bibinfo {volume} {83}},\ \bibinfo {pages} {3174}
  (\bibinfo {year} {1999})}\BibitemShut {NoStop}%
\bibitem [{\citenamefont {Villanueva}\ \emph {et~al.}(2011)\citenamefont
  {Villanueva}, \citenamefont {Karabalin}, \citenamefont {Matheny},
  \citenamefont {Kenig}, \citenamefont {Cross},\ and\ \citenamefont
  {Roukes}}]{villanueva2011}%
  \BibitemOpen
  \bibfield  {author} {\bibinfo {author} {\bibfnamefont {L.~G.}\ \bibnamefont
  {Villanueva}}, \bibinfo {author} {\bibfnamefont {R.~B.}\ \bibnamefont
  {Karabalin}}, \bibinfo {author} {\bibfnamefont {M.~H.}\ \bibnamefont
  {Matheny}}, \bibinfo {author} {\bibfnamefont {E.}~\bibnamefont {Kenig}},
  \bibinfo {author} {\bibfnamefont {M.~C.}\ \bibnamefont {Cross}}, \ and\
  \bibinfo {author} {\bibfnamefont {M.~L.}\ \bibnamefont {Roukes}},\
  }\href@noop {} {\bibfield  {journal} {\bibinfo  {journal} {Nano letters}\
  }\textbf {\bibinfo {volume} {11}},\ \bibinfo {pages} {5054} (\bibinfo {year}
  {2011})}\BibitemShut {NoStop}%
\bibitem [{\citenamefont {Pontin}\ \emph {et~al.}(2014)\citenamefont {Pontin},
  \citenamefont {Bonaldi}, \citenamefont {Borrielli}, \citenamefont
  {Cataliotti}, \citenamefont {Marino}, \citenamefont {Prodi}, \citenamefont
  {Serra},\ and\ \citenamefont {Marin}}]{pontin2014}%
  \BibitemOpen
  \bibfield  {author} {\bibinfo {author} {\bibfnamefont {A.}~\bibnamefont
  {Pontin}}, \bibinfo {author} {\bibfnamefont {M.}~\bibnamefont {Bonaldi}},
  \bibinfo {author} {\bibfnamefont {A.}~\bibnamefont {Borrielli}}, \bibinfo
  {author} {\bibfnamefont {F.}~\bibnamefont {Cataliotti}}, \bibinfo {author}
  {\bibfnamefont {F.}~\bibnamefont {Marino}}, \bibinfo {author} {\bibfnamefont
  {G.}~\bibnamefont {Prodi}}, \bibinfo {author} {\bibfnamefont
  {E.}~\bibnamefont {Serra}}, \ and\ \bibinfo {author} {\bibfnamefont
  {F.}~\bibnamefont {Marin}},\ }\href@noop {} {\bibfield  {journal} {\bibinfo
  {journal} {Physical review letters}\ }\textbf {\bibinfo {volume} {112}},\
  \bibinfo {pages} {023601} (\bibinfo {year} {2014})}\BibitemShut {NoStop}%
\bibitem [{\citenamefont {Sch{\"a}fermeier}\ \emph {et~al.}(2016)\citenamefont
  {Sch{\"a}fermeier}, \citenamefont {Kerdoncuff}, \citenamefont {Hoff},
  \citenamefont {Fu}, \citenamefont {Huck}, \citenamefont {Bilek},
  \citenamefont {Harris}, \citenamefont {Bowen}, \citenamefont {Gehring},\ and\
  \citenamefont {Andersen}}]{schafermeier2016}%
  \BibitemOpen
  \bibfield  {author} {\bibinfo {author} {\bibfnamefont {C.}~\bibnamefont
  {Sch{\"a}fermeier}}, \bibinfo {author} {\bibfnamefont {H.}~\bibnamefont
  {Kerdoncuff}}, \bibinfo {author} {\bibfnamefont {U.~B.}\ \bibnamefont
  {Hoff}}, \bibinfo {author} {\bibfnamefont {H.}~\bibnamefont {Fu}}, \bibinfo
  {author} {\bibfnamefont {A.}~\bibnamefont {Huck}}, \bibinfo {author}
  {\bibfnamefont {J.}~\bibnamefont {Bilek}}, \bibinfo {author} {\bibfnamefont
  {G.~I.}\ \bibnamefont {Harris}}, \bibinfo {author} {\bibfnamefont {W.~P.}\
  \bibnamefont {Bowen}}, \bibinfo {author} {\bibfnamefont {T.}~\bibnamefont
  {Gehring}}, \ and\ \bibinfo {author} {\bibfnamefont {U.~L.}\ \bibnamefont
  {Andersen}},\ }\href@noop {} {\bibfield  {journal} {\bibinfo  {journal}
  {Nature communications}\ }\textbf {\bibinfo {volume} {7}},\ \bibinfo {pages}
  {13628} (\bibinfo {year} {2016})}\BibitemShut {NoStop}%
\bibitem [{\citenamefont {Rossi}\ \emph {et~al.}(2017)\citenamefont {Rossi},
  \citenamefont {Kralj}, \citenamefont {Zippilli}, \citenamefont {Natali},
  \citenamefont {Borrielli}, \citenamefont {Pandraud}, \citenamefont {Serra},
  \citenamefont {Di~Giuseppe},\ and\ \citenamefont {Vitali}}]{rossi2017}%
  \BibitemOpen
  \bibfield  {author} {\bibinfo {author} {\bibfnamefont {M.}~\bibnamefont
  {Rossi}}, \bibinfo {author} {\bibfnamefont {N.}~\bibnamefont {Kralj}},
  \bibinfo {author} {\bibfnamefont {S.}~\bibnamefont {Zippilli}}, \bibinfo
  {author} {\bibfnamefont {R.}~\bibnamefont {Natali}}, \bibinfo {author}
  {\bibfnamefont {A.}~\bibnamefont {Borrielli}}, \bibinfo {author}
  {\bibfnamefont {G.}~\bibnamefont {Pandraud}}, \bibinfo {author}
  {\bibfnamefont {E.}~\bibnamefont {Serra}}, \bibinfo {author} {\bibfnamefont
  {G.}~\bibnamefont {Di~Giuseppe}}, \ and\ \bibinfo {author} {\bibfnamefont
  {D.}~\bibnamefont {Vitali}},\ }\href@noop {} {\bibfield  {journal} {\bibinfo
  {journal} {arXiv:1704.04556}\ } (\bibinfo {year} {2017})}\BibitemShut
  {NoStop}%
\bibitem [{\citenamefont {Patil}\ \emph {et~al.}(2015)\citenamefont {Patil},
  \citenamefont {Chakram}, \citenamefont {Chang},\ and\ \citenamefont
  {Vengalattore}}]{patil2015}%
  \BibitemOpen
  \bibfield  {author} {\bibinfo {author} {\bibfnamefont {Y.}~\bibnamefont
  {Patil}}, \bibinfo {author} {\bibfnamefont {S.}~\bibnamefont {Chakram}},
  \bibinfo {author} {\bibfnamefont {L.}~\bibnamefont {Chang}}, \ and\ \bibinfo
  {author} {\bibfnamefont {M.}~\bibnamefont {Vengalattore}},\ }\href@noop {}
  {\bibfield  {journal} {\bibinfo  {journal} {Physical review letters}\
  }\textbf {\bibinfo {volume} {115}},\ \bibinfo {pages} {017202} (\bibinfo
  {year} {2015})}\BibitemShut {NoStop}%
\bibitem [{\citenamefont {Landau}\ and\ \citenamefont
  {Lifshitz}(2004)}]{landau1976}%
  \BibitemOpen
  \bibfield  {author} {\bibinfo {author} {\bibfnamefont {L.}~\bibnamefont
  {Landau}}\ and\ \bibinfo {author} {\bibfnamefont {E.}~\bibnamefont
  {Lifshitz}},\ }\href@noop {} {\emph {\bibinfo {title} {Mechanics. 3rd}}}\
  (\bibinfo {year} {2004})\BibitemShut {NoStop}%
\bibitem [{\citenamefont {Rugar}\ and\ \citenamefont
  {Gr{\"u}tter}(1991)}]{rugar1991}%
  \BibitemOpen
  \bibfield  {author} {\bibinfo {author} {\bibfnamefont {D.}~\bibnamefont
  {Rugar}}\ and\ \bibinfo {author} {\bibfnamefont {P.}~\bibnamefont
  {Gr{\"u}tter}},\ }\href@noop {} {\bibfield  {journal} {\bibinfo  {journal}
  {Physical Review Letters}\ }\textbf {\bibinfo {volume} {67}},\ \bibinfo
  {pages} {699} (\bibinfo {year} {1991})}\BibitemShut {NoStop}%
\bibitem [{\citenamefont {Turner}\ \emph {et~al.}(1998)\citenamefont {Turner},
  \citenamefont {Miller}, \citenamefont {Hartwell}, \citenamefont {MacDonald}
  \emph {et~al.}}]{turner1998}%
  \BibitemOpen
  \bibfield  {author} {\bibinfo {author} {\bibfnamefont {K.~L.}\ \bibnamefont
  {Turner}}, \bibinfo {author} {\bibfnamefont {S.~A.}\ \bibnamefont {Miller}},
  \bibinfo {author} {\bibfnamefont {P.~G.}\ \bibnamefont {Hartwell}}, \bibinfo
  {author} {\bibfnamefont {N.~C.}\ \bibnamefont {MacDonald}},  \emph {et~al.},\
  }\href@noop {} {\bibfield  {journal} {\bibinfo  {journal} {Nature}\ }\textbf
  {\bibinfo {volume} {396}},\ \bibinfo {pages} {149} (\bibinfo {year}
  {1998})}\BibitemShut {NoStop}%
\bibitem [{\citenamefont {Carr}\ \emph {et~al.}(2000)\citenamefont {Carr},
  \citenamefont {Evoy}, \citenamefont {Sekaric}, \citenamefont {Craighead},\
  and\ \citenamefont {Parpia}}]{carr2000}%
  \BibitemOpen
  \bibfield  {author} {\bibinfo {author} {\bibfnamefont {D.~W.}\ \bibnamefont
  {Carr}}, \bibinfo {author} {\bibfnamefont {S.}~\bibnamefont {Evoy}}, \bibinfo
  {author} {\bibfnamefont {L.}~\bibnamefont {Sekaric}}, \bibinfo {author}
  {\bibfnamefont {H.}~\bibnamefont {Craighead}}, \ and\ \bibinfo {author}
  {\bibfnamefont {J.}~\bibnamefont {Parpia}},\ }\href@noop {} {\bibfield
  {journal} {\bibinfo  {journal} {Applied Physics Letters}\ }\textbf {\bibinfo
  {volume} {77}},\ \bibinfo {pages} {1545} (\bibinfo {year}
  {2000})}\BibitemShut {NoStop}%
\bibitem [{\citenamefont {Rhoads}\ and\ \citenamefont
  {Shaw}(2010)}]{rhoads2010}%
  \BibitemOpen
  \bibfield  {author} {\bibinfo {author} {\bibfnamefont {J.~F.}\ \bibnamefont
  {Rhoads}}\ and\ \bibinfo {author} {\bibfnamefont {S.~W.}\ \bibnamefont
  {Shaw}},\ }\href@noop {} {\bibfield  {journal} {\bibinfo  {journal} {Applied
  Physics Letters}\ }\textbf {\bibinfo {volume} {96}},\ \bibinfo {pages}
  {234101} (\bibinfo {year} {2010})}\BibitemShut {NoStop}%
\bibitem [{\citenamefont {Karabalin}\ \emph {et~al.}(2010)\citenamefont
  {Karabalin}, \citenamefont {Masmanidis},\ and\ \citenamefont
  {Roukes}}]{karabalin2010}%
  \BibitemOpen
  \bibfield  {author} {\bibinfo {author} {\bibfnamefont {R.}~\bibnamefont
  {Karabalin}}, \bibinfo {author} {\bibfnamefont {S.}~\bibnamefont
  {Masmanidis}}, \ and\ \bibinfo {author} {\bibfnamefont {M.}~\bibnamefont
  {Roukes}},\ }\href@noop {} {\bibfield  {journal} {\bibinfo  {journal}
  {Applied Physics Letters}\ }\textbf {\bibinfo {volume} {97}},\ \bibinfo
  {pages} {183101} (\bibinfo {year} {2010})}\BibitemShut {NoStop}%
\bibitem [{\citenamefont {Szorkovszky}\ \emph {et~al.}(2011)\citenamefont
  {Szorkovszky}, \citenamefont {Doherty}, \citenamefont {Harris},\ and\
  \citenamefont {Bowen}}]{s2011}%
  \BibitemOpen
  \bibfield  {author} {\bibinfo {author} {\bibfnamefont {A.}~\bibnamefont
  {Szorkovszky}}, \bibinfo {author} {\bibfnamefont {A.~C.}\ \bibnamefont
  {Doherty}}, \bibinfo {author} {\bibfnamefont {G.~I.}\ \bibnamefont {Harris}},
  \ and\ \bibinfo {author} {\bibfnamefont {W.~P.}\ \bibnamefont {Bowen}},\
  }\href@noop {} {\bibfield  {journal} {\bibinfo  {journal} {Physical review
  letters}\ }\textbf {\bibinfo {volume} {107}},\ \bibinfo {pages} {213603}
  (\bibinfo {year} {2011})}\BibitemShut {NoStop}%
\bibitem [{\citenamefont {Mahboob}\ \emph {et~al.}(2014)\citenamefont
  {Mahboob}, \citenamefont {Okamoto}, \citenamefont {Onomitsu},\ and\
  \citenamefont {Yamaguchi}}]{mahboob2014}%
  \BibitemOpen
  \bibfield  {author} {\bibinfo {author} {\bibfnamefont {I.}~\bibnamefont
  {Mahboob}}, \bibinfo {author} {\bibfnamefont {H.}~\bibnamefont {Okamoto}},
  \bibinfo {author} {\bibfnamefont {K.}~\bibnamefont {Onomitsu}}, \ and\
  \bibinfo {author} {\bibfnamefont {H.}~\bibnamefont {Yamaguchi}},\ }\href@noop
  {} {\bibfield  {journal} {\bibinfo  {journal} {Physical review letters}\
  }\textbf {\bibinfo {volume} {113}},\ \bibinfo {pages} {167203} (\bibinfo
  {year} {2014})}\BibitemShut {NoStop}%
\bibitem [{\citenamefont {Leuch}\ \emph {et~al.}(2016)\citenamefont {Leuch},
  \citenamefont {Papariello}, \citenamefont {Zilberberg}, \citenamefont
  {Degen}, \citenamefont {Chitra},\ and\ \citenamefont {Eichler}}]{leuch2016}%
  \BibitemOpen
  \bibfield  {author} {\bibinfo {author} {\bibfnamefont {A.}~\bibnamefont
  {Leuch}}, \bibinfo {author} {\bibfnamefont {L.}~\bibnamefont {Papariello}},
  \bibinfo {author} {\bibfnamefont {O.}~\bibnamefont {Zilberberg}}, \bibinfo
  {author} {\bibfnamefont {C.~L.}\ \bibnamefont {Degen}}, \bibinfo {author}
  {\bibfnamefont {R.}~\bibnamefont {Chitra}}, \ and\ \bibinfo {author}
  {\bibfnamefont {A.}~\bibnamefont {Eichler}},\ }\href@noop {} {\bibfield
  {journal} {\bibinfo  {journal} {Physical review letters}\ }\textbf {\bibinfo
  {volume} {117}},\ \bibinfo {pages} {214101} (\bibinfo {year}
  {2016})}\BibitemShut {NoStop}%
\bibitem [{\citenamefont {Zwickl}\ \emph {et~al.}(2008)\citenamefont {Zwickl},
  \citenamefont {Shanks}, \citenamefont {Jayich}, \citenamefont {Yang},
  \citenamefont {Bleszynski~Jayich}, \citenamefont {Thompson},\ and\
  \citenamefont {Harris}}]{zwickl2008}%
  \BibitemOpen
  \bibfield  {author} {\bibinfo {author} {\bibfnamefont {B.}~\bibnamefont
  {Zwickl}}, \bibinfo {author} {\bibfnamefont {W.}~\bibnamefont {Shanks}},
  \bibinfo {author} {\bibfnamefont {A.}~\bibnamefont {Jayich}}, \bibinfo
  {author} {\bibfnamefont {C.}~\bibnamefont {Yang}}, \bibinfo {author}
  {\bibfnamefont {A.}~\bibnamefont {Bleszynski~Jayich}}, \bibinfo {author}
  {\bibfnamefont {J.}~\bibnamefont {Thompson}}, \ and\ \bibinfo {author}
  {\bibfnamefont {J.}~\bibnamefont {Harris}},\ }\href@noop {} {\bibfield
  {journal} {\bibinfo  {journal} {Applied Physics Letters}\ }\textbf {\bibinfo
  {volume} {92}},\ \bibinfo {pages} {103125} (\bibinfo {year}
  {2008})}\BibitemShut {NoStop}%
\bibitem [{\citenamefont {J{\"o}ckel}\ \emph {et~al.}(2011)\citenamefont
  {J{\"o}ckel}, \citenamefont {Rakher}, \citenamefont {Korppi}, \citenamefont
  {Camerer}, \citenamefont {Hunger}, \citenamefont {Mader},\ and\ \citenamefont
  {Treutlein}}]{jockel2011}%
  \BibitemOpen
  \bibfield  {author} {\bibinfo {author} {\bibfnamefont {A.}~\bibnamefont
  {J{\"o}ckel}}, \bibinfo {author} {\bibfnamefont {M.~T.}\ \bibnamefont
  {Rakher}}, \bibinfo {author} {\bibfnamefont {M.}~\bibnamefont {Korppi}},
  \bibinfo {author} {\bibfnamefont {S.}~\bibnamefont {Camerer}}, \bibinfo
  {author} {\bibfnamefont {D.}~\bibnamefont {Hunger}}, \bibinfo {author}
  {\bibfnamefont {M.}~\bibnamefont {Mader}}, \ and\ \bibinfo {author}
  {\bibfnamefont {P.}~\bibnamefont {Treutlein}},\ }\href@noop {} {\bibfield
  {journal} {\bibinfo  {journal} {Applied physics letters}\ }\textbf {\bibinfo
  {volume} {99}},\ \bibinfo {pages} {143109} (\bibinfo {year}
  {2011})}\BibitemShut {NoStop}%
\bibitem [{\citenamefont {Wilson-Rae}\ \emph {et~al.}(2011)\citenamefont
  {Wilson-Rae}, \citenamefont {Barton}, \citenamefont {Verbridge},
  \citenamefont {Southworth}, \citenamefont {Ilic}, \citenamefont {Craighead},\
  and\ \citenamefont {Parpia}}]{wilson2011}%
  \BibitemOpen
  \bibfield  {author} {\bibinfo {author} {\bibfnamefont {I.}~\bibnamefont
  {Wilson-Rae}}, \bibinfo {author} {\bibfnamefont {R.}~\bibnamefont {Barton}},
  \bibinfo {author} {\bibfnamefont {S.}~\bibnamefont {Verbridge}}, \bibinfo
  {author} {\bibfnamefont {D.}~\bibnamefont {Southworth}}, \bibinfo {author}
  {\bibfnamefont {B.}~\bibnamefont {Ilic}}, \bibinfo {author} {\bibfnamefont
  {H.}~\bibnamefont {Craighead}}, \ and\ \bibinfo {author} {\bibfnamefont
  {J.}~\bibnamefont {Parpia}},\ }\href@noop {} {\bibfield  {journal} {\bibinfo
  {journal} {Physical review letters}\ }\textbf {\bibinfo {volume} {106}},\
  \bibinfo {pages} {047205} (\bibinfo {year} {2011})}\BibitemShut {NoStop}%
\bibitem [{\citenamefont {Yu}\ \emph {et~al.}(2012)\citenamefont {Yu},
  \citenamefont {Purdy},\ and\ \citenamefont {Regal}}]{yu2012}%
  \BibitemOpen
  \bibfield  {author} {\bibinfo {author} {\bibfnamefont {P.-L.}\ \bibnamefont
  {Yu}}, \bibinfo {author} {\bibfnamefont {T.}~\bibnamefont {Purdy}}, \ and\
  \bibinfo {author} {\bibfnamefont {C.}~\bibnamefont {Regal}},\ }\href@noop {}
  {\bibfield  {journal} {\bibinfo  {journal} {Physical review letters}\
  }\textbf {\bibinfo {volume} {108}},\ \bibinfo {pages} {083603} (\bibinfo
  {year} {2012})}\BibitemShut {NoStop}%
\bibitem [{\citenamefont {Villanueva}\ and\ \citenamefont
  {Schmid}(2014)}]{villanueva2014}%
  \BibitemOpen
  \bibfield  {author} {\bibinfo {author} {\bibfnamefont {L.~G.}\ \bibnamefont
  {Villanueva}}\ and\ \bibinfo {author} {\bibfnamefont {S.}~\bibnamefont
  {Schmid}},\ }\href@noop {} {\bibfield  {journal} {\bibinfo  {journal}
  {Physical review letters}\ }\textbf {\bibinfo {volume} {113}},\ \bibinfo
  {pages} {227201} (\bibinfo {year} {2014})}\BibitemShut {NoStop}%
\bibitem [{\citenamefont {Li}\ \emph {et~al.}(2016)\citenamefont {Li},
  \citenamefont {Zhang}, \citenamefont {You}, \citenamefont {Li},\ and\
  \citenamefont {Peng}}]{li2016}%
  \BibitemOpen
  \bibfield  {author} {\bibinfo {author} {\bibfnamefont {Z.}~\bibnamefont
  {Li}}, \bibinfo {author} {\bibfnamefont {Q.}~\bibnamefont {Zhang}}, \bibinfo
  {author} {\bibfnamefont {X.}~\bibnamefont {You}}, \bibinfo {author}
  {\bibfnamefont {Y.}~\bibnamefont {Li}}, \ and\ \bibinfo {author}
  {\bibfnamefont {K.}~\bibnamefont {Peng}},\ }\href@noop {} {\bibfield
  {journal} {\bibinfo  {journal} {Applied Physics Letters}\ }\textbf {\bibinfo
  {volume} {109}},\ \bibinfo {pages} {191903} (\bibinfo {year}
  {2016})}\BibitemShut {NoStop}%
\bibitem [{\citenamefont {Chakram}\ \emph {et~al.}(2014)\citenamefont
  {Chakram}, \citenamefont {Patil}, \citenamefont {Chang},\ and\ \citenamefont
  {Vengalattore}}]{chakram2014}%
  \BibitemOpen
  \bibfield  {author} {\bibinfo {author} {\bibfnamefont {S.}~\bibnamefont
  {Chakram}}, \bibinfo {author} {\bibfnamefont {Y.}~\bibnamefont {Patil}},
  \bibinfo {author} {\bibfnamefont {L.}~\bibnamefont {Chang}}, \ and\ \bibinfo
  {author} {\bibfnamefont {M.}~\bibnamefont {Vengalattore}},\ }\href@noop {}
  {\bibfield  {journal} {\bibinfo  {journal} {Physical review letters}\
  }\textbf {\bibinfo {volume} {112}},\ \bibinfo {pages} {127201} (\bibinfo
  {year} {2014})}\BibitemShut {NoStop}%
\bibitem [{\citenamefont {Reinhardt}\ \emph {et~al.}(2016)\citenamefont
  {Reinhardt}, \citenamefont {M{\"u}ller}, \citenamefont {Bourassa},\ and\
  \citenamefont {Sankey}}]{reinhardt2016}%
  \BibitemOpen
  \bibfield  {author} {\bibinfo {author} {\bibfnamefont {C.}~\bibnamefont
  {Reinhardt}}, \bibinfo {author} {\bibfnamefont {T.}~\bibnamefont
  {M{\"u}ller}}, \bibinfo {author} {\bibfnamefont {A.}~\bibnamefont
  {Bourassa}}, \ and\ \bibinfo {author} {\bibfnamefont {J.~C.}\ \bibnamefont
  {Sankey}},\ }\href@noop {} {\bibfield  {journal} {\bibinfo  {journal}
  {Physical Review X}\ }\textbf {\bibinfo {volume} {6}},\ \bibinfo {pages}
  {021001} (\bibinfo {year} {2016})}\BibitemShut {NoStop}%
\bibitem [{\citenamefont {Norte}\ \emph {et~al.}(2016)\citenamefont {Norte},
  \citenamefont {Moura},\ and\ \citenamefont {Gr{\"o}blacher}}]{norte2016}%
  \BibitemOpen
  \bibfield  {author} {\bibinfo {author} {\bibfnamefont {R.~A.}\ \bibnamefont
  {Norte}}, \bibinfo {author} {\bibfnamefont {J.~P.}\ \bibnamefont {Moura}}, \
  and\ \bibinfo {author} {\bibfnamefont {S.}~\bibnamefont {Gr{\"o}blacher}},\
  }\href@noop {} {\bibfield  {journal} {\bibinfo  {journal} {Physical review
  letters}\ }\textbf {\bibinfo {volume} {116}},\ \bibinfo {pages} {147202}
  (\bibinfo {year} {2016})}\BibitemShut {NoStop}%
\bibitem [{\citenamefont {Gavartin}\ \emph {et~al.}(2013)\citenamefont
  {Gavartin}, \citenamefont {Verlot},\ and\ \citenamefont
  {Kippenberg}}]{gavartin2013}%
  \BibitemOpen
  \bibfield  {author} {\bibinfo {author} {\bibfnamefont {E.}~\bibnamefont
  {Gavartin}}, \bibinfo {author} {\bibfnamefont {P.}~\bibnamefont {Verlot}}, \
  and\ \bibinfo {author} {\bibfnamefont {T.~J.}\ \bibnamefont {Kippenberg}},\
  }\href@noop {} {\bibfield  {journal} {\bibinfo  {journal} {Nature
  Communications}\ }\textbf {\bibinfo {volume} {4}} (\bibinfo {year}
  {2013})}\BibitemShut {NoStop}%
\bibitem [{\citenamefont {Lifshitz}\ and\ \citenamefont
  {Cross}(2010)}]{lifshitz2010}%
  \BibitemOpen
  \bibfield  {author} {\bibinfo {author} {\bibfnamefont {R.}~\bibnamefont
  {Lifshitz}}\ and\ \bibinfo {author} {\bibfnamefont {M.}~\bibnamefont
  {Cross}},\ }\href@noop {} {\emph {\bibinfo {title} {Nonlinear dynamics of
  nanomechanical resonators}}}\ (\bibinfo  {publisher} {Weinheim, Germany:
  Wiley-VCH Verlag GmbH \& Co. KGaA},\ \bibinfo {year} {2010})\ pp.\ \bibinfo
  {pages} {221--266}\BibitemShut {NoStop}%
\bibitem [{\citenamefont {Ma}\ \emph {et~al.}(2005)\citenamefont {Ma},
  \citenamefont {Ye}, \citenamefont {Wei},\ and\ \citenamefont
  {Zhang}}]{ma2005}%
  \BibitemOpen
  \bibfield  {author} {\bibinfo {author} {\bibfnamefont {H.}~\bibnamefont
  {Ma}}, \bibinfo {author} {\bibfnamefont {C.}~\bibnamefont {Ye}}, \bibinfo
  {author} {\bibfnamefont {D.}~\bibnamefont {Wei}}, \ and\ \bibinfo {author}
  {\bibfnamefont {J.}~\bibnamefont {Zhang}},\ }\href@noop {} {\bibfield
  {journal} {\bibinfo  {journal} {Physical review letters}\ }\textbf {\bibinfo
  {volume} {95}},\ \bibinfo {pages} {233601} (\bibinfo {year}
  {2005})}\BibitemShut {NoStop}%
\bibitem [{\citenamefont {Sheng}\ \emph {et~al.}(2012)\citenamefont {Sheng},
  \citenamefont {Khadka},\ and\ \citenamefont {Xiao}}]{sheng2012}%
  \BibitemOpen
  \bibfield  {author} {\bibinfo {author} {\bibfnamefont {J.}~\bibnamefont
  {Sheng}}, \bibinfo {author} {\bibfnamefont {U.}~\bibnamefont {Khadka}}, \
  and\ \bibinfo {author} {\bibfnamefont {M.}~\bibnamefont {Xiao}},\ }\href@noop
  {} {\bibfield  {journal} {\bibinfo  {journal} {Physical review letters}\
  }\textbf {\bibinfo {volume} {109}},\ \bibinfo {pages} {223906} (\bibinfo
  {year} {2012})}\BibitemShut {NoStop}%
\bibitem [{\citenamefont {Khadka}\ \emph {et~al.}(2013)\citenamefont {Khadka},
  \citenamefont {Sheng},\ and\ \citenamefont {Xiao}}]{khadka2013}%
  \BibitemOpen
  \bibfield  {author} {\bibinfo {author} {\bibfnamefont {U.}~\bibnamefont
  {Khadka}}, \bibinfo {author} {\bibfnamefont {J.}~\bibnamefont {Sheng}}, \
  and\ \bibinfo {author} {\bibfnamefont {M.}~\bibnamefont {Xiao}},\ }\href@noop
  {} {\bibfield  {journal} {\bibinfo  {journal} {Physical review letters}\
  }\textbf {\bibinfo {volume} {111}},\ \bibinfo {pages} {223601} (\bibinfo
  {year} {2013})}\BibitemShut {NoStop}%
\bibitem [{\citenamefont {Fiore}\ \emph {et~al.}(2011)\citenamefont {Fiore},
  \citenamefont {Yang}, \citenamefont {Kuzyk}, \citenamefont {Barbour},
  \citenamefont {Tian},\ and\ \citenamefont {Wang}}]{fiore2011}%
  \BibitemOpen
  \bibfield  {author} {\bibinfo {author} {\bibfnamefont {V.}~\bibnamefont
  {Fiore}}, \bibinfo {author} {\bibfnamefont {Y.}~\bibnamefont {Yang}},
  \bibinfo {author} {\bibfnamefont {M.~C.}\ \bibnamefont {Kuzyk}}, \bibinfo
  {author} {\bibfnamefont {R.}~\bibnamefont {Barbour}}, \bibinfo {author}
  {\bibfnamefont {L.}~\bibnamefont {Tian}}, \ and\ \bibinfo {author}
  {\bibfnamefont {H.}~\bibnamefont {Wang}},\ }\href@noop {} {\bibfield
  {journal} {\bibinfo  {journal} {Physical review letters}\ }\textbf {\bibinfo
  {volume} {107}},\ \bibinfo {pages} {133601} (\bibinfo {year}
  {2011})}\BibitemShut {NoStop}%
\bibitem [{\citenamefont {Mahboob}\ and\ \citenamefont
  {Yamaguchi}(2008)}]{mahboob2008}%
  \BibitemOpen
  \bibfield  {author} {\bibinfo {author} {\bibfnamefont {I.}~\bibnamefont
  {Mahboob}}\ and\ \bibinfo {author} {\bibfnamefont {H.}~\bibnamefont
  {Yamaguchi}},\ }\href@noop {} {\bibfield  {journal} {\bibinfo  {journal}
  {Nature nanotechnology}\ }\textbf {\bibinfo {volume} {3}},\ \bibinfo {pages}
  {275} (\bibinfo {year} {2008})}\BibitemShut {NoStop}%
\end{thebibliography}
%

\end{document}